\documentclass[10pt,twoside,twocolumn]{pnas-new}
\usepackage{float}
\usepackage{hyperref}
\usepackage{xcolor}
\usepackage{hyperref}
\usepackage{natbib}
\usepackage{subfiles}
\usepackage{aas_macros}
\usepackage{amssymb}
\usepackage{amsmath}
\usepackage{deluxetable} 
\usepackage{cuted}

\usepackage{graphicx}
\usepackage{dcolumn}
\usepackage{bm}
\usepackage{cases}
\usepackage{caption}
\usepackage{subcaption}

\newcommand{\be}{\begin{equation}}
\newcommand{\ee}{\end{equation}}
\newcommand{\JADEStwelve}{JADES-GS-z12-0~}
\newcommand{\JADESeleven}{JADES-GS-z11-0~}

\newcommand{\JADESzthirteen}{JADES-GS-z13-0~}
\newcommand{\JADESfz}{JADES-GS-z14-0~}
\newcommand{\JADESfo}{JADES-GS-z14-1~}

\newcommand{\unit}[1]{\mathrm{#1}}
\newcommand{\GeV}{\unit{GeV}}
\newcommand{\cc}{\unit{cm}^3}
\newcommand{\percc}{\unit{cm}^{-3}}

\newcommand{\Msun}{M_{\odot}}

\newcommand{\sigmav}{\langle\sigma v\rangle}

\templatetype{pnasresearcharticle} 

\title{Spectroscopic Supermassive Dark Star candidates}

\author[a,1]{Cosmin Ilie}
\author[a]{Sayed Shafaat Mahmud}
\author[b]{Jillian Paulin}
\author[c,d,e,1]{Katherine Freese}

\affil[a]{ Department of Physics and Astronomy, Colgate University,
13 Oak Dr., Hamilton, NY 13346, U.S.A.
}
\affil[b]{ Department of Physics and Astronomy, University of Pennsylvania,
209 South 33rd Street,  Philadelphia, PA 19104-6396, U.S.A.
}
\affil[c]{Weinberg Institute for Theoretical Physics, Texas Center for Cosmology and Astroparticle Physics,
Department of Physics, University of Texas, Austin, TX 78712, USA;}
\affil[d]{Department of Physics, Stockholm University, Stockholm, Sweden}
\affil[e]{Nordic Institute for Theoretical Physics (NORDITA), Stockholm, Sweden}

\leadauthor{Ilie}

 \significancestatement{In 2007 Spolyar, Freese, and Gondolo proposed the idea of Dark Stars. Some of the first stars in the Universe might have been powered by dark matter annihilations rather than nuclear fusion. They could form out of pristine Hydrogen and Helium clouds at the centers of protogalaxies, where there is a sufficient abundance of dark matter to serve as their heat source. They are very bright diffuse puffy objects and grow to be very massive. In fact, in 2023 three of us identified the first three photometric Dark Star candidates in the JWST NIRCam data. In this paper we use JWST NIRSpec data and identify four spectroscopic Dark Star candidates, one of them being the most distant object ever observed.}

\authorcontributions{C.I., J.P., S.M., and K.F. designed research; C.I., S.M., and JP. performed research; C.I., S.M., J.P., and K.F. wrote the manuscript }
\authordeclaration{The authors declare no competing interests}
\correspondingauthor{\textsuperscript{1}To whom correspondence should be addressed: \href{mailto:cilie@colgate.edu}{cilie@colgate.edu} or \href{mailto:ktfreese@utexas.edu}{ktfreese@utexas.edu}}

\keywords{Cosmology $|$ Dark Matter $|$ First Stars $|$ Dark Stars}
	
\begin{abstract}
Dark Stars, i.e. early stars composed almost entirely of hydrogen and helium but powered by Dark Matter, could form 
in zero metallicity clouds located close to the center of high redshift Dark Matter halos. In 2023 three of us identified (in a PNAS work) the first three photometric Dark Star candidates: \JADESeleven, \JADEStwelve, and \JADESzthirteen. We report here our results of a followup analysis based on available NIRSpec JWST data. We find that \JADESeleven and \JADESzthirteen are spectroscopically consistent with a Dark Star interpretation. Moreover, we find two additional spectroscopic Dark Star candidates: \JADESfz and \JADESfo, with the former being the most distant luminous object ever observed. We furthermore identify a feature in its spectrum indicative of the smoking gun signature of Dark Stars: the He~II $\lambda$1640 absorption line. In view ALMA’s recent identification of a probable O~III nebular emission line in the spectrum of \JADESfz, the simple interpretation of this object as an isolated Dark Star is unlikely. If both spectral features survive follow-up observations it would imply a Dark Star embedded in a metal rich environment, requiring theoretical refinements of the formation of evolution of Dark Stars, which in previous studies were assumed to form in isolation, without any companions.   
\end{abstract}

\dates{This manuscript was compiled on \today}
\doi{\url{www.pnas.org/cgi/doi/10.1073/pnas.XXXXXXXXXX}}

\begin{document}

\maketitle
\thispagestyle{firststyle}
\ifthenelse{\boolean{shortarticle}}{\ifthenelse{\boolean{singlecolumn}}{\abscontentformatted}{\abscontent}}{}


The first stars to form in the Universe, when it was roughly 200 million years old~\citep[e.g.][]{Abel:2001, Barkana:2000, Bromm:2003, Yoshida:2006, OShea:2007, Yoshida:2008, Bromm:2009}, may have been Dark Stars (DSs),
composed almost entirely of hydrogen and helium from the Big Bang but powered by Dark Matter (DM) annihilation rather than by nuclear fusion~\citep{Spolyar:2008dark,Freese:2008ds, Iocco:2008}.
 Although dark matter constitutes only $< 0.1\%$ of the stellar mass, this amount is sufficient to power the star for millions to billions of years.
 The relevant types of dark matter for heating the stars include Weakly Interacting Massive Particles (WIMPs)~\footnote{This is the case for most previous works on Dark Stars}, and Self Interacting Dark Matter (SIDM)~\citep{Wu:2022wzw}. 
Starting from their inception at $\sim 1 M_\odot$, Dark Stars accrete mass from their surroundings to become supermassive stars, 
some even reaching masses $\geq 10^6 M_\odot$ and luminosities $10^9 L_\odot$, making them detectable with the James Webb Space Telescope (JWST)~\citep{Freese:2010smds,Zackrisson:2010HighZDS,Ilie:2012} or the upcoming Roman Space Telescope (RST)~\citep{Zhang:2022}.
In Ref.~\cite{Ilie:2023JADES}, three of us found several SMDS candidates in the JADES survey of the JWST: \JADESeleven, \JADEStwelve, and \JADESzthirteen. 
These objects are good fits to photometric NIRCam data predicted for SMDS using the \texttt{TLUSTY}~\citep{hubeny2017tlusty} code for estimating SEDs. 
One Super Massive Dark Star (SMDS) can be as bright as an entire early galaxy of stars.
As shown by Ref.~\cite{Zhang:2022}, it is difficult, yet not impossible, to differentiate between isolated SMDSs and certain types of early galaxies based on spectroscopic or photometric data; spectroscopic data over a wide range of wavelengths (i.e. NIRSpec \& ALMA) should help with this, as we will discuss.

The methodology of our current paper provides several improvements over our previous work~\citep{Ilie:2023JADES}:  
1) We compute more accurate spectra and estimates of the size of SMDS (as would be seen by JWST); 2) we use new spectroscopic data from NIRSpec (rather than
the NIRCam photometry we used previously); and 3) we perform more careful statistical analyses in fitting the theoretical predictions to the data.  Most interestingly,
we find two new SDMS candidates in JWST data. 
Specifically, using publicly available NIRSpec data, we find that the measured spectra of two of the three previously identified photometric candidates (\JADESzthirteen and \JADESeleven) are consistent with a Dark Star interpretation (the third object we previously examined has some metal lines in it that imply 
it is not an isolated dark star). Moreover, we identify two more spectroscopic Dark Star candidates: \JADESfo and \JADESfz, with the former being the most distant spectroscopically confirmed object ever observed. Furthermore, we find a tentative ($S/N\sim 2.4$)  He~II$\lambda$1640 absorption feature (a smoking gun for Dark Stars) in the NIRSpec spectrum for \JADESfz. As we were wrapping up this manuscript Ref.~\citep{carniani2025eventful}  reported  evidence of an O~III nebular emission line in the ALMA measured spectrum of \JADESfz. If both spectral features are confirmed, this object cannot be an isolated Dark Star but rather may be a Dark Star embedded in a metal rich environment as we plan to study in future work (see Discussion \& Conclusion).

Before proceeding, we wish to point out that the brightest and most massive SMDS could resolve several conundrums of high redshift JWST data. 
First, JWST has discovered a larger than expected number of very bright high redshift objects~\citep{GLASSz13, Maisies:2022, z16.CEERS93316:2022, z17.Schrodinger:2022,JADES:2022a,JADES:2022b,Labbe:2022,RedMonsters:2024}. The high stellar masses inferred for the most luminous of these objects, if they are galaxies,
requires an inexplicably high efficiency rate of conversion of gas to stellar material. \footnote{Explanations have been put forward that alleviate this tension, e.g. due to top heavy Initial Mass Functions~\citep[e.g.][]{Finkelstein_2023,harikane2023purespectroscopicconstraintsuv,Yung:2024,JZ:2025} or feedback free star burst formation episodes~\citep{Dekel_2023}, to name a few. }  
We argue that some of these overly massive and bright objects could instead be SMDSs.

In addition, SMDSs may provide an explanation of the many early Supermassive Black Holes (SMBHs) that are
otherwise to hard to explain.  
Once the SMDS runs out of dark matter fuel, it collapses to a black hole, e.g. of $10^6 M_\odot$, thus providing excellent seeds for the even larger SMBHs found already at early times. JWST has found many such enormous SMBHs.  Black Holes (BHs) at $z\sim 6$ appear to be about three orders of magnitude more massive than what the local relation between BH and stellar masses would imply~\citep{BHdorm24}. The most striking such object is UHZ-1, an enormous Supermassive Black Hole (SMBH) at $z\sim 10$, with a mass of about ten billion suns. Emitting light when the universe was a mere 500 Myr old, UHZ-1 is considered to be the first indication of the need for very massive Black Hole seeds~\citep{Bogdan:2023UHZ1}.  UHZ1 is just the tip of the iceberg when it comes  $z\gtrsim 6$ quasars observed harboring SMBHs that are too big to have been seeded by regular stars~\citep[e.g.][]{Wang:2019,Inayoshi:2020,Lupi:2021}. Even the most massive zero metallicity nuclear burning stars (i.e. $M\sim 10^3\Msun$)\footnote{For references on zero metallicity, a.k.a. Pop~III, stars see, for example Refs.~\cite{Abel:2001,Barkana:2000,Bromm:2003,Yoshida:2006,OShea:2007,Yoshida:2008,Bromm:2009}} could have seeded such gigantic SMBHs only if they accreted for long periods of time at super Eddington rates. More likely, the data implies the need for heavy Black Hole seeds, at high redshifts~\citep{Bogdan:2023UHZ1,ilie2023uhz1}. Direct Collapse Black Holes (DCBHs)~\citep[e.g.][]{Loeb:1994wv,Belgman:2006,Lodato:2006hw,Natarajan:2017,barrow:2018,Whalen:2020,Inayoshi:2020}  and Dark Stars~\citep[e.g.][]{Spolyar:2008dark,Freese:2010smds, Banik:2019,Singh_2023,cammelli2024formationsupermassiveblackholes}~\footnote{Note that Dark Stars are sometimes called Pop~III.1 stars in the literature~\citep[e.g.][]{Banik:2019,Singh_2023,cammelli2024formationsupermassiveblackholes}} are two proposed mechanisms for producing such heavy BH seeds in the first few hundred million years after the Big Bang. 

Thus Dark Stars are a plausible solution to two of the most pressing puzzles in Astronomy today: (i). the many brighter than expected high-z compact objects found by JWST and (ii). the origin of the massive BH seeds required to explain many of the most distant quasars observed.  We note that the case studied here is that of Dark Stars powered solely by the  DM  passing through the central region of the  halo  (rather than additional DM 
that can be captured via elastic scattering with nuclei in the star, leading to hotter DSs).
Further, for simplicity we assume the DM is composed of 100 GeV WIMPs with a canonical annihilation cross section, but argue that our results are relatively unchanged for a variety of DM masses and cross sections.
The rest of the paper is organized as follows: we begin by briefly discussing Dark Stars (DS) and reviewing their main properties, we then proceed to explain our methodology of identifying spectroscopic Dark Star candidates. The paper ends with our Results, followed by a Discussion and Conclusions sections.

\section*{Dark Stars} In view of the lack of direct observations, numerical simulations are still the primary tool to describe the properties of the first stars~\citep[e.g.][]{Abel:2001, Barkana:2000, Bromm:2003, Yoshida:2006, OShea:2007, Yoshida:2008, Bromm:2009}, and galaxies~\citep[e.g.][]{Loeb:2010, Bromm:2011, Gnedin:2016, Dayal:2018, Yung:2019, Behroozi:2020} in the Universe. The first stars formed roughly 100-400 Myrs after the Big Bang ($z \sim 20-10$) as a consequence of the gravitational collapse of pristine, zero metallicity, molecular hydrogen  clouds at the center of $10^6-10^8 M_\odot$ minihaloes. If the role of Dark Matter heating can be neglected during the collapse of the protostellar hydrogen clouds, the outcome is the formation of nuclear burning Population~III stars. Those grow via mass accretion, reaching masses of (at most) $10^3\Msun$~\cite[e.g.][]{Hirano:2014}, and populate the first small galaxies.  However, as shown by Ref.~\cite{Spolyar:2008dark}, if Dark Matter is formed of Weakly Interacting Massive Particles (WIMPs), then DM heating inside a proto-stellar gas cloud at the center of high-z DM micro-halos can be non-negligible and it can even halt the gravitational collapse well before the central temperatures reach the critical values needed for nuclear fusion to begin. As such, a different type of first star can form: a Dark Star~\citep{Spolyar:2008dark,Freese:2008ds,Spolyar:2009}.~\footnote{For a review of Dark Stars see Ref.~\cite{Freese:2016dark}.} 

Those exotic objects are made primarily of H and He, with Dark Matter accounting for less than $0.1\%$ of their mass. Since DM annihilations are much more efficient at converting mass to energy than nuclear fusion, this small amount of DM can power the star, while maintaining a relatively cool core (i.e. $T_c\lesssim 10^7$~K). Those cool ($T_{eff}\lesssim 10^4$~K) puffy ($R_{DS}\sim 10$~AU) stars can accrete the surrounding baryons almost indefinitely, since the radiative feedback effects that shut off accretion in the case of Pop~III stars are not significant for Dark Stars~\citep{Freese:2010smds}. As such, they can grow to become supermassive (SMDSs), reaching masses $\sim 10^6 M_\odot$ and luminosities $\sim 10^{10} L_\odot$, making them visible to JWST~\cite{Freese:2010smds,Ilie:2012,Ilie:2023JADES} or Roman Space Telescope (RST)~\citep{Zhang:2022}. In view of HSTs sensitivity and the fact that no plausible Dark Star candidates were identified with it, it is expected that isolated DS very seldom survive past $z\lesssim 10$~\citep{Ilie:2012}.

Below we review some of the basic particle physics ingredients needed for the formation of Dark Stars. We considered two types of DM particles:  Weakly Interacting Dark Matter (WIMPs, in most papers) and Self Interacting Dark Matter (SIDM) ~\citep{Wu:2022wzw}.~\footnote{As it turns out, for both WIMPs and SIDM the emerging Dark Stars have very similar observable properties~\citep{Wu:2022wzw}}
The energy production per unit volume provided by the annihilation of two DM particles is:  
\begin{equation}\label{eq:heat}
    Q = m_\chi n_\chi^2 \langle \sigma v\rangle = \langle \sigma v\rangle \rho_\chi^2 / m_\chi,
\end{equation}
 where $m_\chi \sim 1 {\rm GeV} - 10 {\rm TeV}$ is the DM mass, $n_\chi$ is the DM number density, $\rho_\chi$ is the DM energy density. We assumed the standard annihilation cross section (the value that produces the correct DM abundance in the Universe today): $\langle \sigma v\rangle = 3 \times 10^{-26} {\rm cm}^3/{\rm s}$, although cross sections several orders of magnitude smaller or larger would work equally well since a decrease in $\sigmav$ can be easily compensated by an increase in $\rho_\chi^2/m_\chi$. The three conditions necessary for the formation of Dark Stars are: 1) sufficient DM density, 2) DM annihilation products become trapped inside the star, and 3) the DM heating rate beats the cooling rate of the collapsing cloud. Those are easily met for the case of WIMPs~\citep{Spolyar:2008dark} and SIDM~\citep{Wu:2022wzw}.  

The high DM density condition can be met in one of two ways. As molecular clouds collapse inside high-z minihaloes, they pull in more Dark Matter as well. This is a purely gravitational effect, which leads to the significant enhancement of Dark Matter densities at the centers of said halos. Using adiabatic contraction (AC)~\cite{Blumenthal:1985}~\footnote{A formalism confirmed in more sophisticated analyses in the context of Dark Stars in Ref.~\citep{Freese:2008dmdens}} one can estimate that the local DM densities where the first stars form can be as high as $\gtrsim 10^{14}~\GeV\percc$. Since many DM particles are on chaotic or box orbits, the central DM density can be replenished and kept high for millions (to billions) of years.  If or when this adiabatically contracted DM reservoir is depleted, the star begins to collapse. As a result it will reach sufficiently high baryonic densities, such  that it is able to capture further DM particles from its vicinity via elastic scattering of the DM with the atoms in the star.  Dark Stars powered by Captured DM are somewhat hotter than those powered via AC, since they underwent a contraction phase. For simplicity, in this work we only consider Dark Stars powered by AC DM. Even with this restricting assumption, we find that all four objects analyzed are consistent with a Dark Star interpretation. 

We use stellar structure solving codes to build up the mass of DSs from an initial value of $\sim 1 M_\odot$ via accretion to potentially very large masses.  We simulate Dark Star evolutionary tracks with two different types of stellar codes:  one of which assumes that the DS can be approximated as polytrope of variable index (i.e. transitioning from 3/2 in the small mass stages, to 3 once the DS becomes fully radiation pressure dominated), and the \texttt{MESA} stellar evolution code~\citep{Paxton2011}.  In both cases, we find essentially the same results~\cite{Rindler-Daller:2015SMDS}. For simplicity, in this work we only consider 100 GeV WIMPs powering Dark Stars, and use the polytropic equilibrium models obtained in Ref.~\cite{Freese:2010smds}. From the equilibrium structure we can derive observable properties, such as flux, and use that to identify candidates, as explained in detail in the next section. 

In previous work, we obtained spectra for DSs using {\texttt{TLUSTY}}.  We assumed that DSs could be treated as point sources in the data as they would be unresolved by JWST. However, in this paper we make an important improvement.  
SEDs are  refined using the \texttt{CLOUDY} code~\citep{chatzikos20232023}, which accounts for possible nebular emission generated by the ionization of surrounding hydrogen by Supermassive Dark Stars. This effect has been first considered by Ref.~\cite{Zackrisson:2011} for Supermassive Dark Stars formed via DM Capture.  For Dark Stars formed via AC (such as those studied by us here), we find that the continuum spectra relevant to JWST remain largely unmodified, whereas nebular emission lines are present at wavelengths larger than studied in the JWST filters. However, Dark Stars powering a nebula are no longer point sources and may be extended.

\section*{Method} We use the most recent publicly available JWST Advanced Deep Extragalactic Survey (JADES) NIRSpec data in order to identify spectroscopic Dark Star candidates. A first cut-off is imposed on the data, as follows: we select objects with confirmed spectroscopic redshift $z_{spec}\gtrsim 10$, which are either extremely compact or consistent with a point source, and for which no metal emission lines have been conclusively identified.~\footnote{Note that in view of the results of Ref.~\cite{GS12:2024jades}, who identify carbon nebular emission in JADES NIRSpec data for \JADEStwelve, this object can no longer be considered as an isolated Dark Star candidate. It could, instead be a Dark Star embedded within a Pop~III/II galaxy, and we plan to test this hypothesis in future work.} We specifically focus on the NIRSpec observations of JADES-GS-z11 and JADES-GS-z13, as described by~\cite{hainline2024searching}, and JADES-GS-z14-0 and JADES-GS-z14-1, as reported by~\cite{carniani2024shining}. These observations do not reveal any confident spectral lines with a signal-to-noise ratio exceeding 3 (within the JADES data; see below for remarks on ALMA data). As a result, we prioritize spectral fitting over the identification of individual spectral lines in our analysis.

For those objects that are consistent with a point source in the JWST data, (such as \JADESfo in this work), we model them as pure Dark Stars without surrounding nebulae, i.e. we assume almost no gas left around them to be ionized. From the evolutionary tracks derived using polytropes in Ref.~\cite{Freese:2010smds} (see Fig.~1 and Table.~3 there) we extract relevant parameters (i.e. stellar mass, stellar radius, and surface temperature) needed to calculate the spectral energy distribution (SED) for Dark Stars. We do so using \texttt{TLUSTY}~\citep{Hubeney:1988,hubeny2017tlusty}, and assume, as appropriate for Dark Stars, zero metallicity stellar atmospheres and a primordial (i.e. Big Bang Nucleosynthesis)  H and He abundance. It is important to note that fixing $m_\chi$ (as we did) leads to the restframe SED being dependent essentially only on one parameter, the stellar mass. As such, we have only two free parameters with which we model high-z point source objects as pure dark stars: redshift (z) and stellar mass ($M_{DS}$) for which we find optimal values in order to fit the JADES spectral data, as described in more detail later in this section. 

 We model the other three objects considered, which are compact but for which a half-radius (or angular size) is reported in the discovery papers, as Dark Stars powering a spherical hydrogen nebula.  Therefore, the pure Dark Star SEDs are further refined using the \texttt{CLOUDY} code~\citep{chatzikos20232023}, which accounts for possible nebular emission generated by the ionization of surrounding hydrogen by Supermassive Dark Stars. While Dark Stars are not hot enough to turn off mass accretion via feedback effects, the most massive ones can power a nebula, as first shown by Ref.~\cite{Zackrisson:2011}. As such, for those objects we have three free parameters that determine observable spectra: redshift (z), stellar mass ($M_{DS}$), and the H density of the nebular cloud ($n_H$).  Remarkably, we demonstrate that even with this minimal set of parameters, we can achieve compelling Dark Star fits to all JADES galaxy candidates considered, as seen in Fig.~\ref{fig:SEDFits} (bottom panel).

 As default, we set the gravitational lensing factor to be $\mu = 1$ (no magnification).~\footnote{In our previous paper, we allowed $\mu$ to vary as a proxy for a finer grid of DS mass, since we previously had runs for limited DS masses; we found that our fits to data were always optimized for $\mu$ near 1.} The only case with nonzero $\mu$ is that of \JADESfz, for which the discovery paper determined $\mu = 1.2$.
 
{\bf Spectral Fitting Technique:} In order to fit our Dark Star models to NIRSpec JADES data, we employ the Nelder-Mead algorithm, a robust optimization technique, to minimize the mean squared error between the simulated and observed spectral data. Observational uncertainties are incorporated through Monte Carlo simulations. For each point in the observed spectra, we define a range of values based on its upper and lower uncertainty bounds. A Gaussian probability distribution is generated for each spectral point, and we draw 100 random samples from these distributions. This approach allows us to perform 100 independent fits of the SED to the observed spectra, resulting in a range of parameter estimates rather than single-point predictions. Consequently, our analysis yields a comprehensive understanding of the likely values for the stellar mass, redshift, and where appropriate H density ($n_H$) of the nebula, while accounting for the uncertainties inherent in the observational data. In our MC runs we bias the value for the prior of $n_H$ within a range that would lead to a Str\"{o}mgren radius (i.e. radius of the HII nebular region surrounding the star) consistent with the reported physical size of the object itself. Our current approach is clearly superior to the simplistic $\chi^2$ analysis we 
used in our first paper to identify DS candidates in JWST.

{\bf Radial Profile Fitting Technique:} Three of the objects considered by us in this work (i.e. \JADESeleven, \JADESzthirteen, \JADESfz) are extremely compact, with radii of at most a few hundred pc, whereas \JADESfo is consistent with a point source.  We model the first three as Dark Stars surrounded by a spherical hydrogen nebula. Their radial profiles are simulated using the JWST Exposure Time Calculator (ETC) Pandeia~\citep{Pickering_2016} and compared against radial profiles (in the standard F200W NIRCam band) reported in the literature~\citep{hainline2024searching,carniani2024shining}. We account for the uncertainties in the measured radial profiles via a Monte Carlo (MC) simulation, as explained below. The flux emitted by each object (which we input from our best fit Dark Star spectral models described above), along with the angular size ($a$) of the object and the S\'{e}rsic index ($n$)~\footnote{The S\'{e}rsic profile~\citep{Sersic:1968} is a commonly assumed empirical law describing how the intensity falls with distance from the center of galaxies or star clusters. Namely, $I\sim \exp {(- R^{1/n})}$, with $n$ being the S\'{e}rsic index. } are the three main ingredients needed to simulate radial profiles for our Dark Stars surrounded by a nebula. For the flux, we use the best fit obtained via the procedure described above, whereas the other two are free parameters we fit for using our MC simulations. Specifically, for each case, a normal distribution is generated for the observed radial profile points, using the reported uncertainties. We sample points from this distribution 75 times, and find best fit for our two free parameters ($a$ and $n$) using the standard Nedler-Mead optimization, which needs initial guesses for the free parameters. Since we assumed a spherical nebula, we bias the choice of our prior of the S\'{e}rsic index ($n$) to be sampled from a normal distribution centered around $n=4$~\citep{Sersic:1968}. For $a$, our initial guess is informed by the value of the angular size for each object, as reported in Refs.~\cite{hainline2024searching, carniani2024shining}. We find that our Dark Star models have a radial profile that fits well the reported JWST data (see Fig.~\ref{fig:morphology}). 

\section*{Results} Below we show that \JADESeleven, \JADESzthirteen, \JADESfz, and \JADESfo can each be modeled by Dark Stars, via the procedure described in our Methods section.

\begin{figure*}[!htb]
\centering
\begin{subfigure}{0.45\textwidth}
    \includegraphics[width=\textwidth]{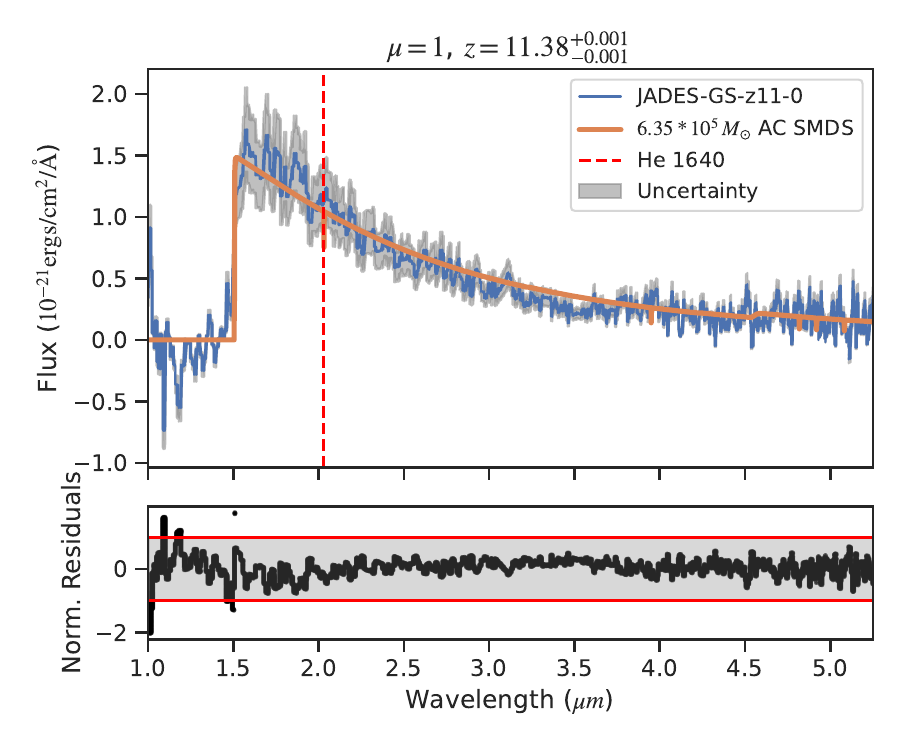}
    \caption{Spectra and DS fit for \JADESeleven}
    \label{fig:SEDa}
\end{subfigure}
\hfill
\begin{subfigure}{0.45\textwidth}
    \includegraphics[width=\textwidth]{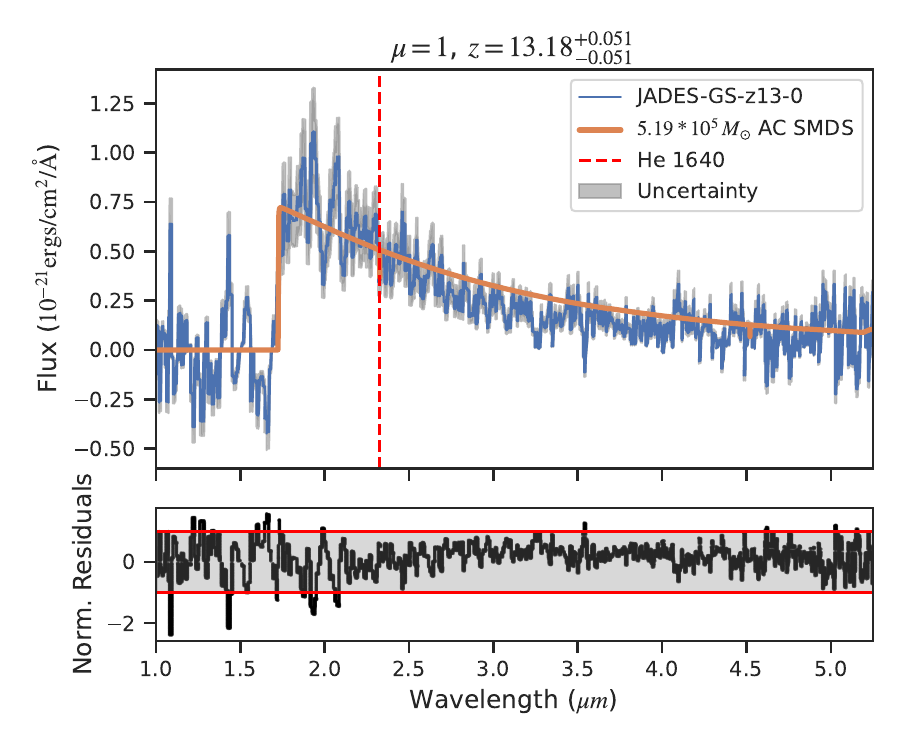}
    \caption{Spectra and DS fit for \JADESzthirteen}
    \label{fig:SEDb}
\end{subfigure}

\begin{subfigure}{0.45\textwidth}
    \includegraphics[width=\textwidth]{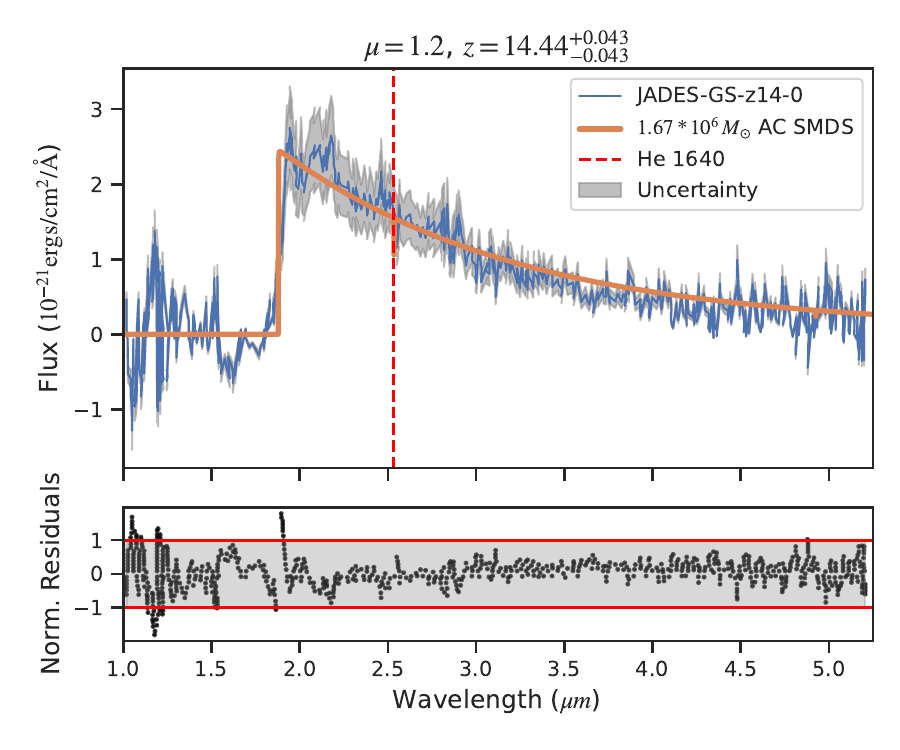}
    \caption{Spectra and DS fit for \JADESfz}
    \label{fig:SEDc}
\end{subfigure}
\hfill
 \begin{subfigure}{0.45\textwidth}
       \includegraphics[width=\textwidth]{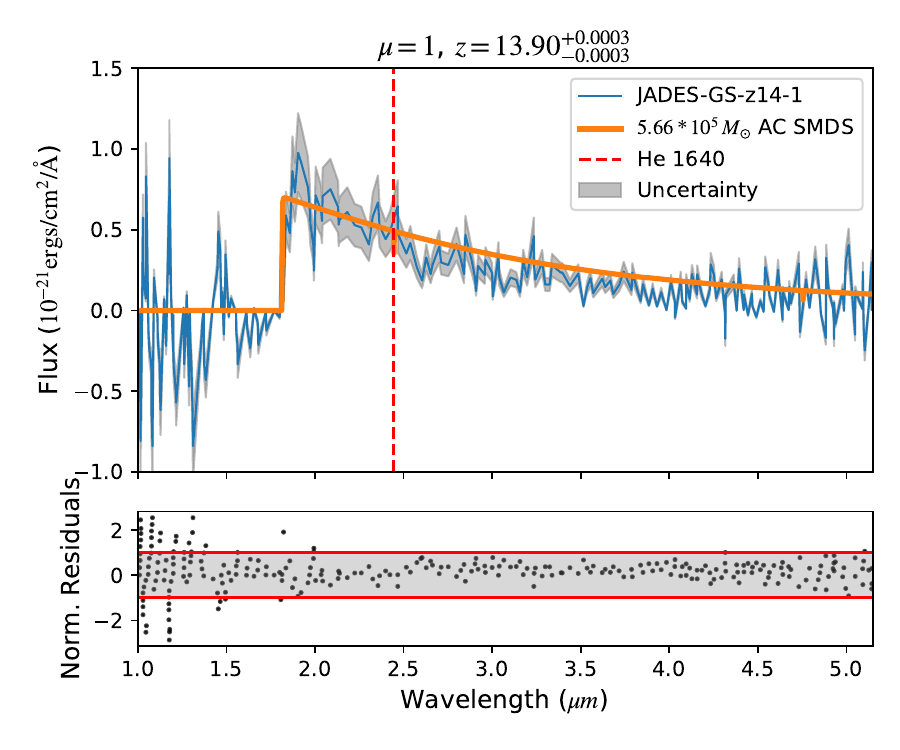}
    \caption{Spectra and DS fit for \JADESfo}
    \label{fig:SEDd}
\end{subfigure}

\caption{The first four Supermassive Dark Star spectral candidates. The data (blue line) and uncertainty band (shaded gray) plotted against our best fit Dark Star models (orange line). The red dashed line represents where He~II$\lambda-1640$ absorption feature, expected only for Dark Stars, might be observed. 
The normalized residuals ($\frac{F_{simul}-F_{measured}} {\sigma_{measured}}$) displayed in the lower panels  of each of the SEDs show that our Supermassive Dark Star models lie consistently within 1-$\sigma$ of the NIRSpec data for each of the four objects considered. The vertical drop in the flux of the SMDS represents the Lyman break, as expected for $z\gtrsim 6$ objects due to absorption by neutral H along the line of sight~\citep{Gunn-Peterson:1965}. The best fit redshift, with associated uncertainty, can be In the title of each plot we display the values assumed for the gravitational lensing factor ($\mu$) and the best fit values for $z_{spec}$ (along with uncertainties).}
\label{fig:SEDFits}
\end{figure*}

\begin{figure*}[!htb]
\centering
\begin{subfigure}{0.45\textwidth}
    \includegraphics[width=\textwidth]{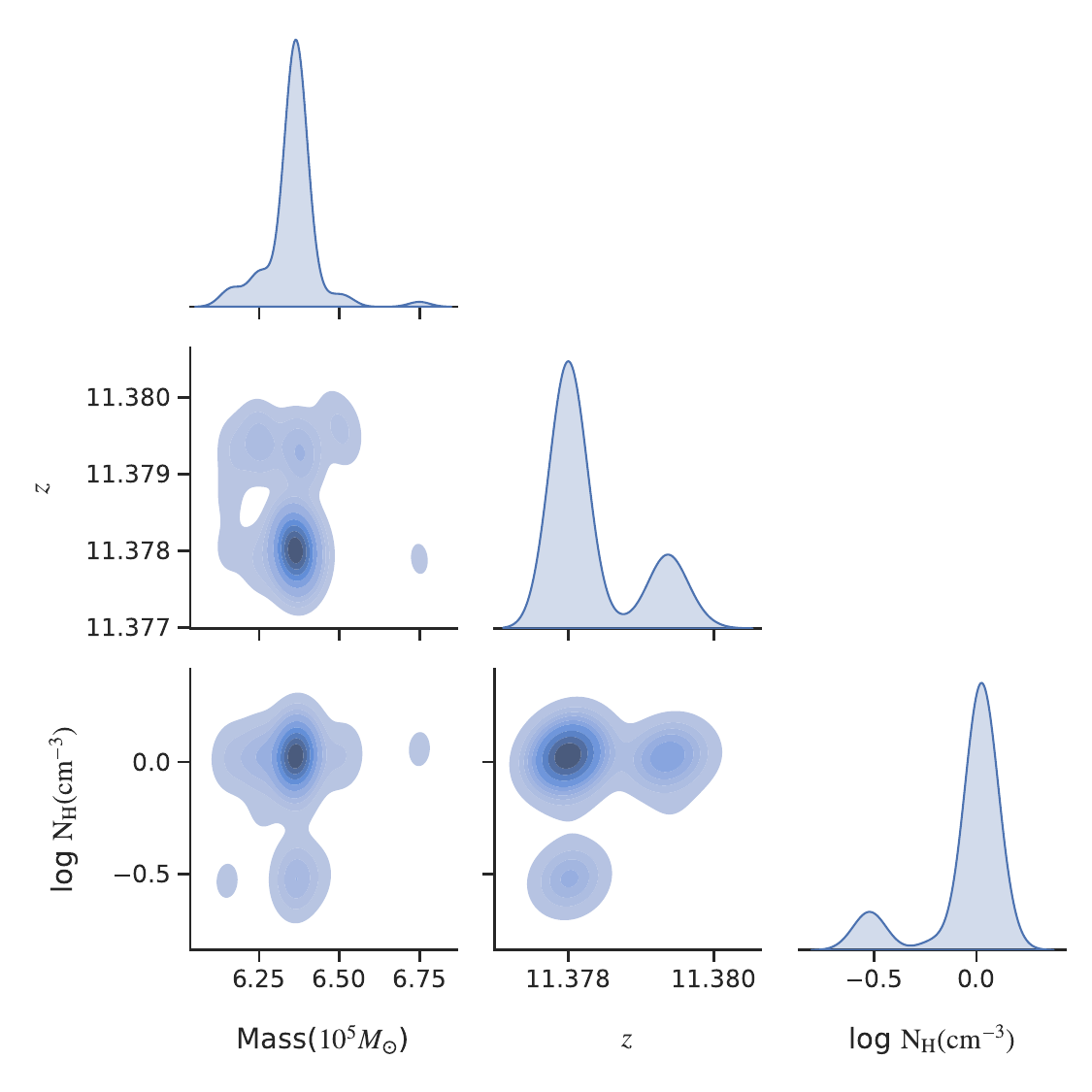}
    \caption{SED MC posteriors for \JADESeleven}
    \label{fig:PostSED_a}
\end{subfigure}
\hfill
\begin{subfigure}{0.45\textwidth}
    \includegraphics[width=\textwidth]{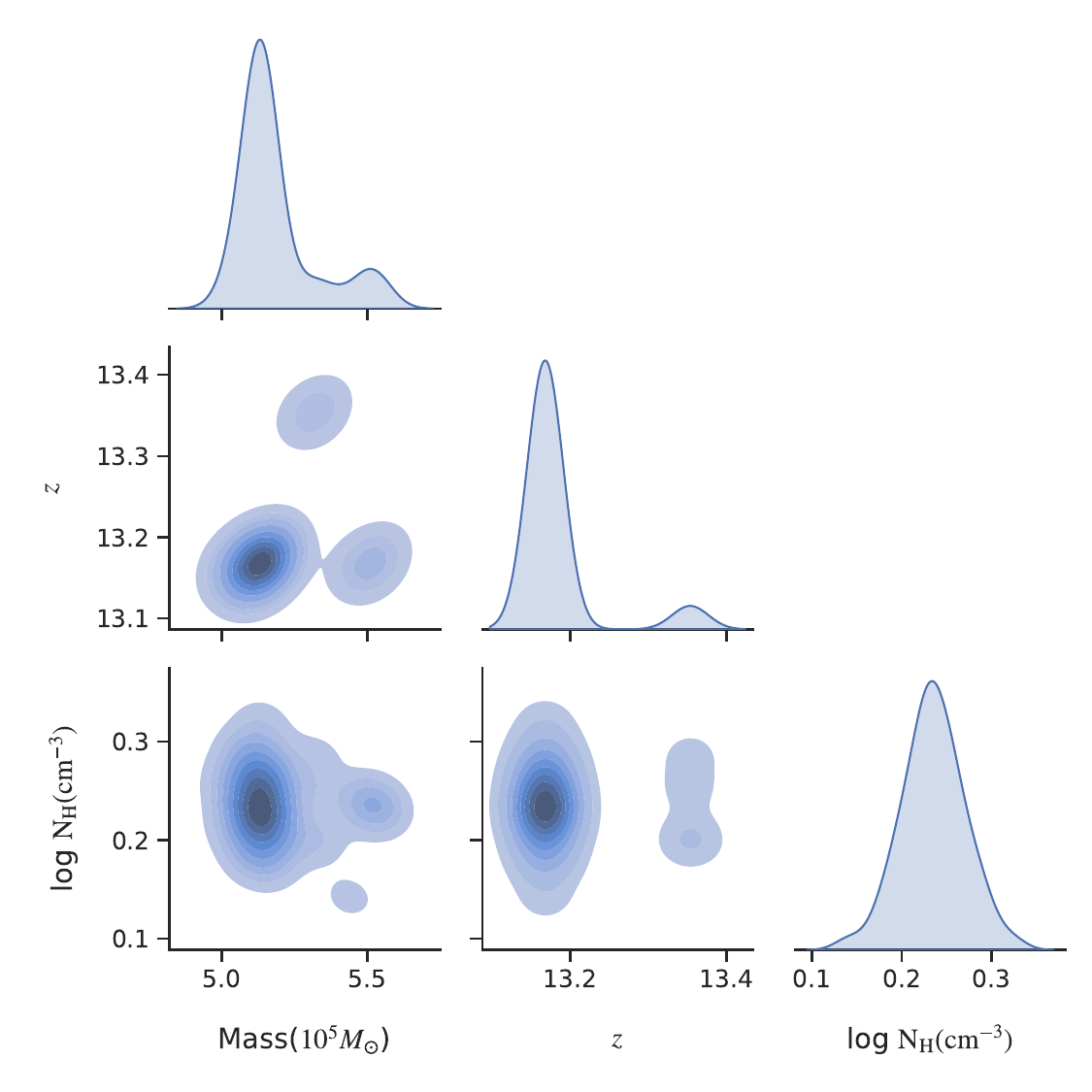}
    \caption{SED MC posteriors for \JADESzthirteen}
    \label{fig:PostSED_b}
\end{subfigure}

\begin{subfigure}{0.45\textwidth}
    \includegraphics[width=\textwidth]{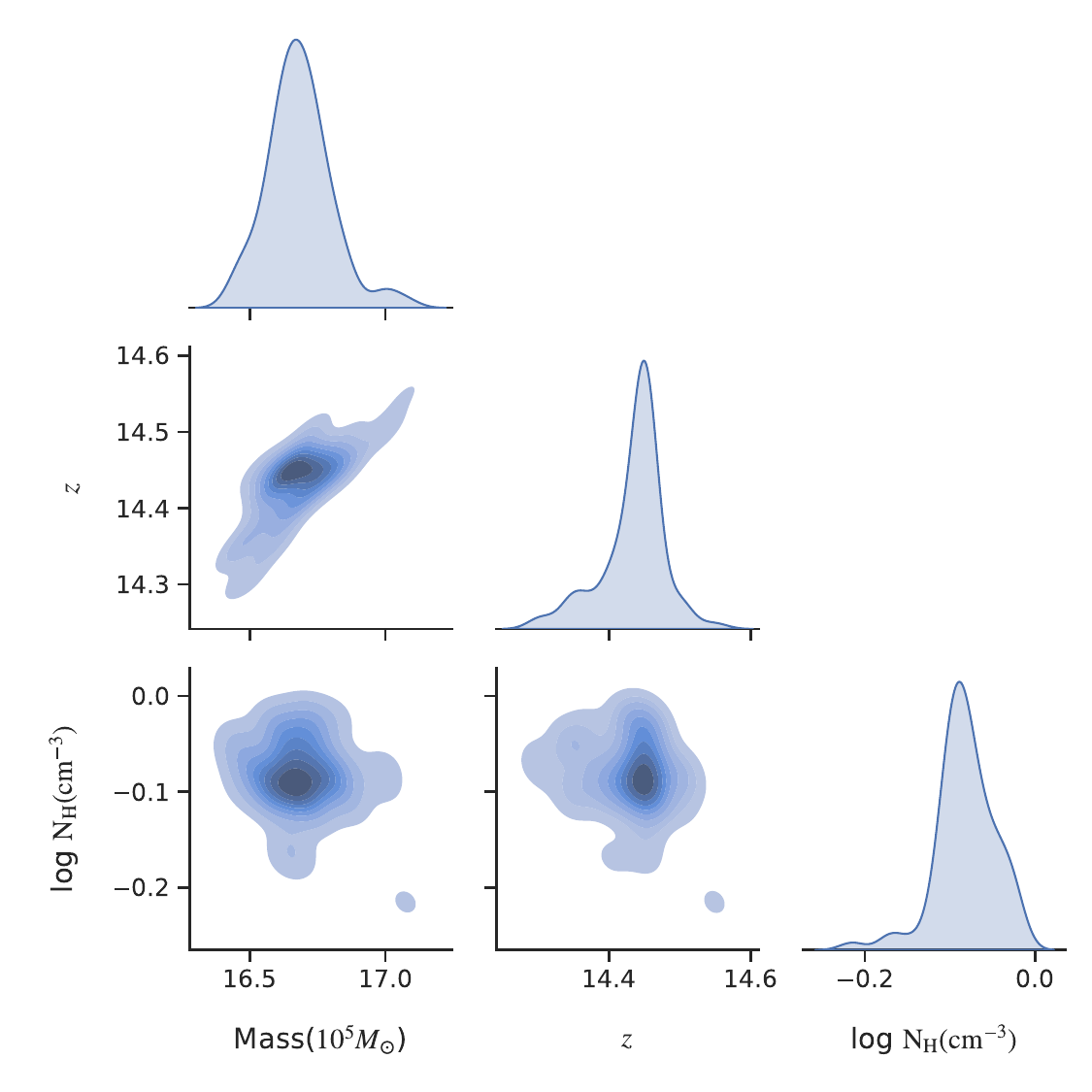}
    \caption{SED MC posteriors for \JADESfz}
    \label{fig:PostSED_c}
\end{subfigure}
\hfill
 \begin{subfigure}{0.45\textwidth}
       \includegraphics[width=\textwidth]{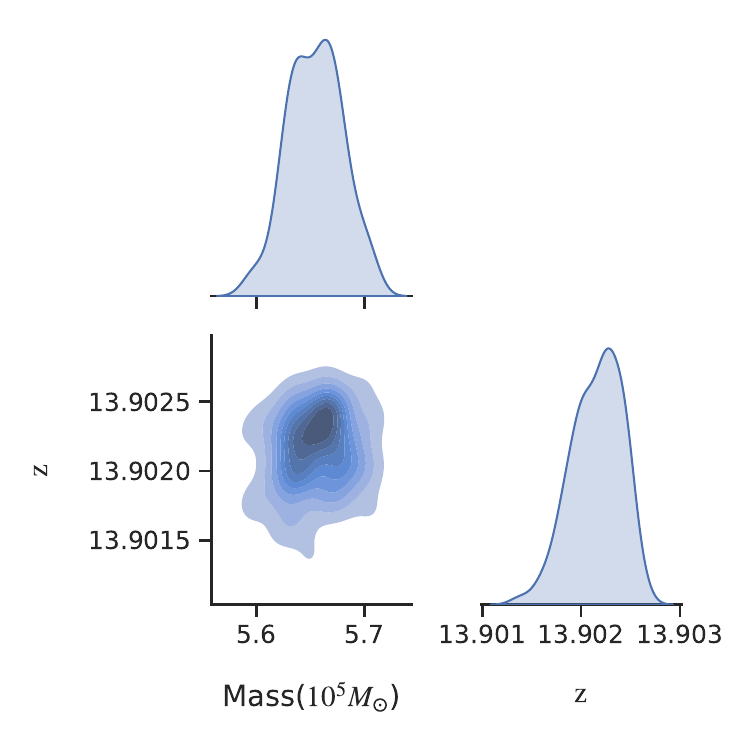}
    \caption{SMDS SED fit for \JADESfo}
    \label{fig:PostSED_d}
    \end{subfigure}

\caption{Corner plots representing 2D and 1D marginalized posteriors for the free parameters of our Supermassive Dark Star spectral fits (as described in Methods). Note that for \JADESfo (lower right hand plot), since it is not resolved, we only need stellar mass ($M$) and redshift ($z$), whereas for the other three objects we include the number density of hydrogen ($n_H$) in the nebular cloud surrounding the Dark Star. We assumed SMDSs are powered by 100 GeV WIMPs adiabatically contracted (AC) inside the star.}
\label{fig:PosteriorsSED}
\end{figure*}

{\bf Spectral Fits:} In Figs.~\ref{fig:SEDFits} and~\ref{fig:PosteriorsSED} we plot our best fit Dark Star models (obtained according to the method described in the previous section) against JWST data~\citep[from][]{hainline2024searching,carniani2024shining} for the four objects considered. Specifically, in Fig.~\ref{fig:SEDFits} we plot our best fit Supermassive Dark Star models against JWST JADES NIRSpec data for all four objects, whereas Fig.~\ref{fig:PosteriorsSED} presents corner plots for the 2D and 1D marginalized posteriors for the free parameters of our SMDSs models.

The normalized residuals displayed at the bottom of each of the SEDs from Fig.~\ref{fig:SEDFits} show that our Supermassive Dark Star models lie consistently within 1-$\sigma$ of the NIRSpec data for each of the four objects considered. Therefore, at the level the continuum of the emission flux they are each consistent with a Dark Star interpretation. No conclusive emission lines are detected with NIRSpec for \JADESzthirteen, \JADESfz~\footnote{ALMA followup observations for this object seem to favor the existence of an [O~III]88$\mu$m emission line~\citep{carniani2025eventful}. We discuss this aspect further at the end of this section.} or \JADESfo, thus as of now those objects are consistent with both a Dark Star and a galaxy interpretation. Only deeper spectra will make the disambiguation possible. In the future, any evidence of metal lines would rule out our simplistic model of an isolated SMDS (or a SMDS powering a zero metallicity nebula), whereas a detection of a He~II $\lambda1640$ absorption feature would confirm the Dark Star hypothesis. For \JADESeleven NIRSpec PRISM data hints  of three metal lines: the [O~II]$\lambda\lambda$3726, 3729$\AA$ doublet and the [N~III]$\lambda$3869$\AA$~\citep{hainline2024searching}. The higher resolution NIRSpec G395M grating spectrum shows no evidence of the [O~II]$\lambda$3726$\AA$ part of the [O~II] doublet, leading to an unphysically high [O~II]$\lambda$3729 / [O~II]$\lambda$3726 observed flux ratio~\citep{hainline2024searching}. In view of this, and the relatively low SNR of the two remaining putative lines (i.e. $SNR=2-3$) we consider \JADESeleven a valid SMDSs candidate (see also Fig.~\ref{fig:SEDa}).

In Table~\ref{tab:jades_properties}(pg.~\pageref{tab:jades_properties}) we summarize the best fit parameters (including 1-$\sigma$ uncertainties) of our Supermassive Dark Star models for the four spectroscopic candidates identified in this work. One can see that the best fit mass for all of the candidates studied here is of order $\sim 10^6 M_\odot$. ~\footnote{Note that in order to save space, in Table~\ref{tab:jades_properties} we use the short names for each object, i.e. we omit the JADES designation.} 

\begin{table}
\centering
\caption{ Summary of SMDSs best fit parameters for the four spectroscopic candidates identified, One can see that the best fit mass for all of the candidates studied here is of order $\sim 10^6 M_\odot$. \label{tab:jades_properties}}
\begin{tabular}{llll}
\toprule
Object Name & Mass ($10^5$ $M_\odot$) & Redshift ($z$) & $\log (\mathrm{N_{H}/\cc)}$ \\
\midrule
GS-z11 & $6.35^{+0.08}_{-0.04}$  & $11.38^{+0.18}_{-0.04}$ & $0.59^{+0.18}_{-0.04}$ \\
GS-z13 & $5.19^{+0.13}_{-0.13}$ & $13.18^{+0.24}_{-0.05}$ & $0.23^{+0.03}_{-0.03}$ \\
GS-z14-0 & $16.68^{+0.12}_{-0.08}$ &$14.43^{+0.04}_{-0.04}$ & $0.03^{+0.03}_{-0.03}$ \\
 GS-z14-1 & $5.65^{+0.02}_{-0.02}$ & $13.90^{+0.0003}_{-0.0003}$ & N/A\\
\bottomrule
\end{tabular}
\end{table}

\begin{figure*}[!htb]
\centering
\begin{subfigure}{0.3\textwidth}
    \includegraphics[width=\textwidth]{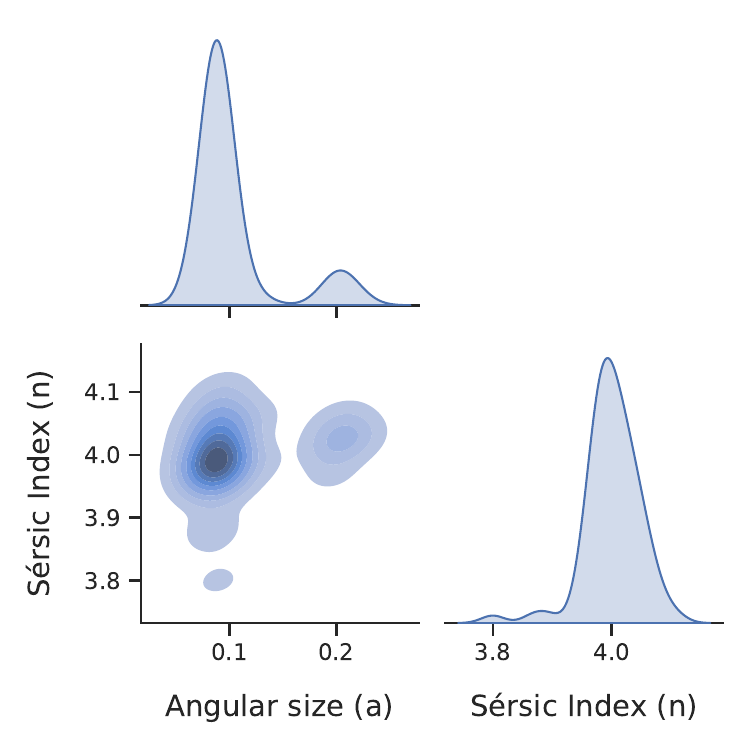}
    \caption{Morphology MC posteriors for \JADESeleven}
    \label{fig:PostMorph_a}
\end{subfigure}
\hfill
\begin{subfigure}{0.3\textwidth}
    \includegraphics[width=\textwidth]{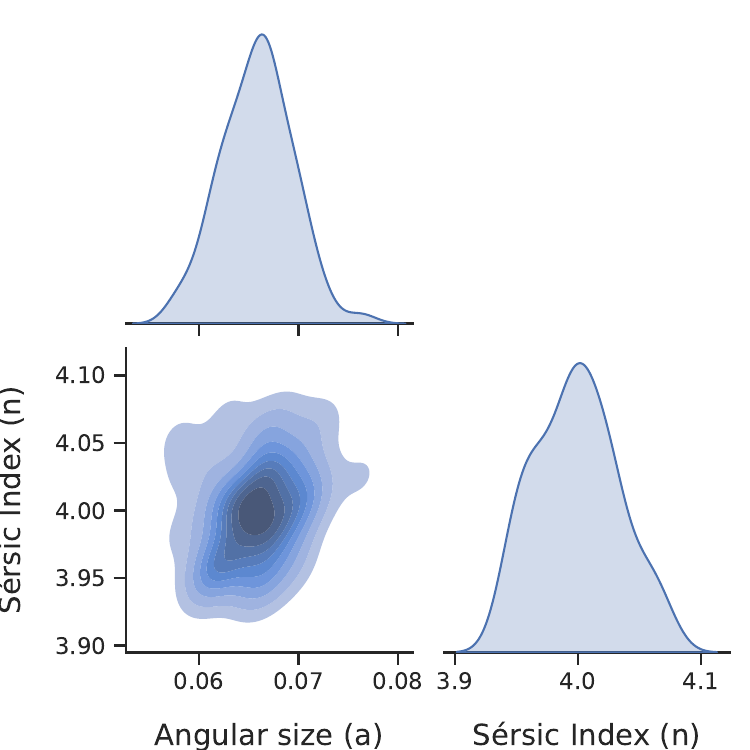}
    \caption{Morphology MC posteriors for \JADESzthirteen}
    \label{fig:PostMorph_b}
\end{subfigure}
\hfill
\begin{subfigure}{0.3\textwidth}
    \includegraphics[width=\textwidth]{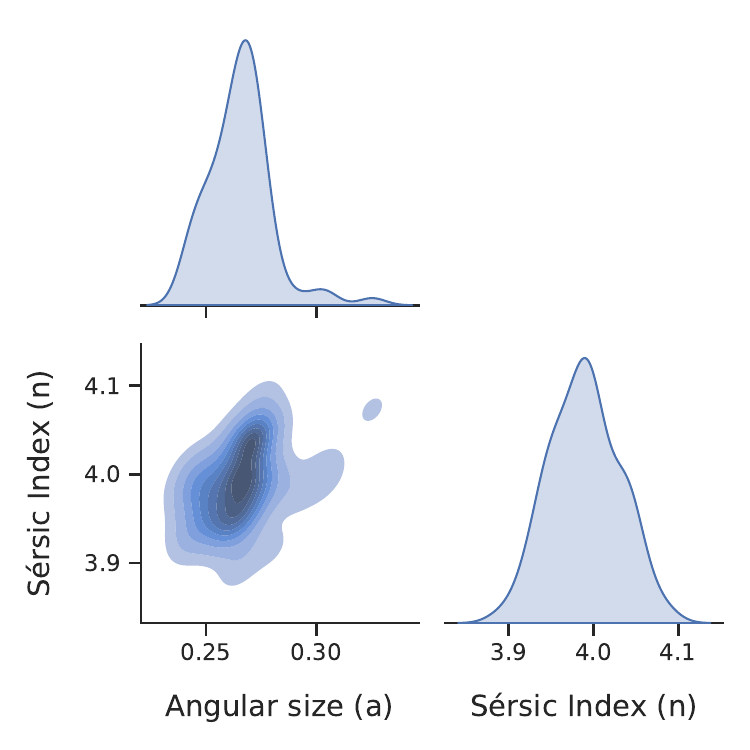}
    \caption{Morphology MC posteriors for \JADESfz}
    \label{fig:PostMorph_c}
\end{subfigure}

 \begin{subfigure}{0.3\textwidth}
       \includegraphics[width=\textwidth]{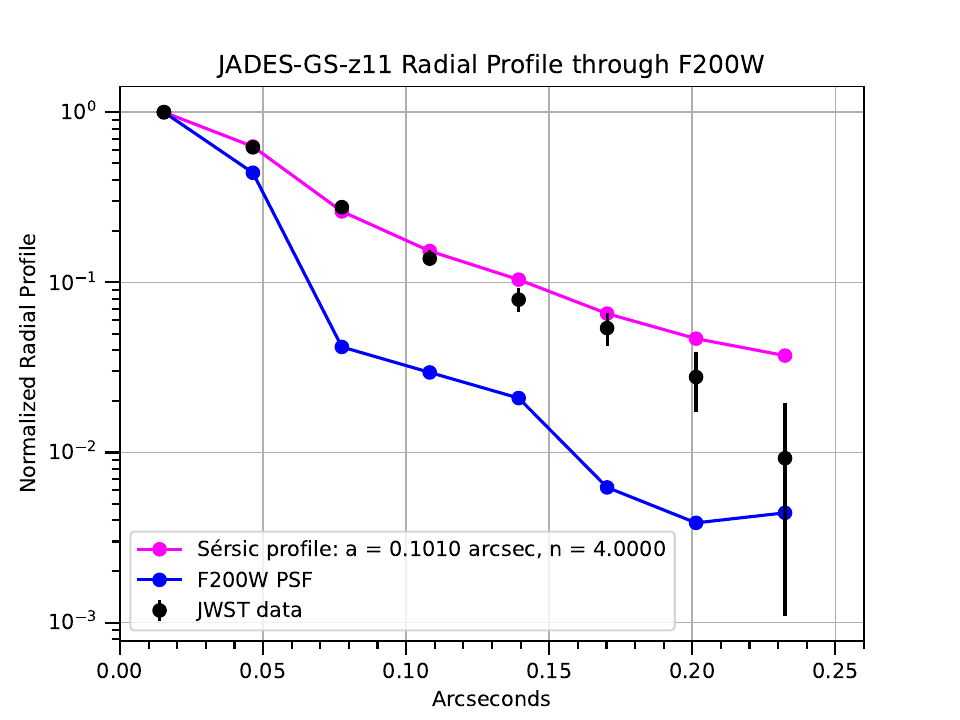}
    \caption{Morphology for \JADESeleven}
    \label{fig:Morph_d}
    \end{subfigure}
\hfill
\begin{subfigure}{0.3\textwidth}
       \includegraphics[width=\textwidth]{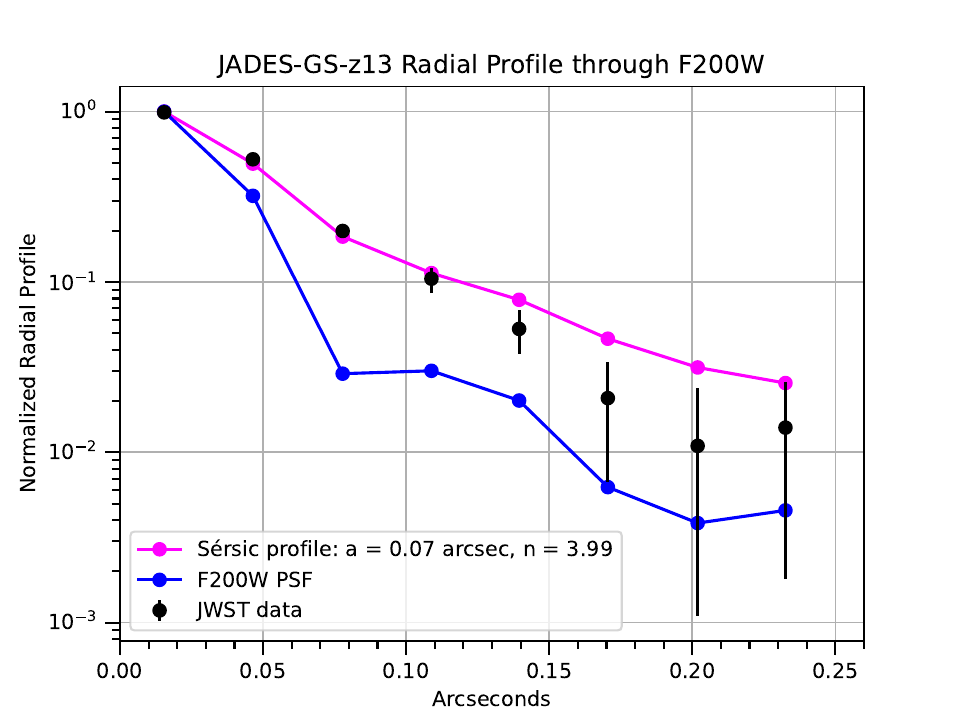}
    \caption{Morphology for \JADESzthirteen}
    \label{fig:Morph_e}
    \end{subfigure}
\hfill
\begin{subfigure}{0.3\textwidth}
       \includegraphics[width=\textwidth]{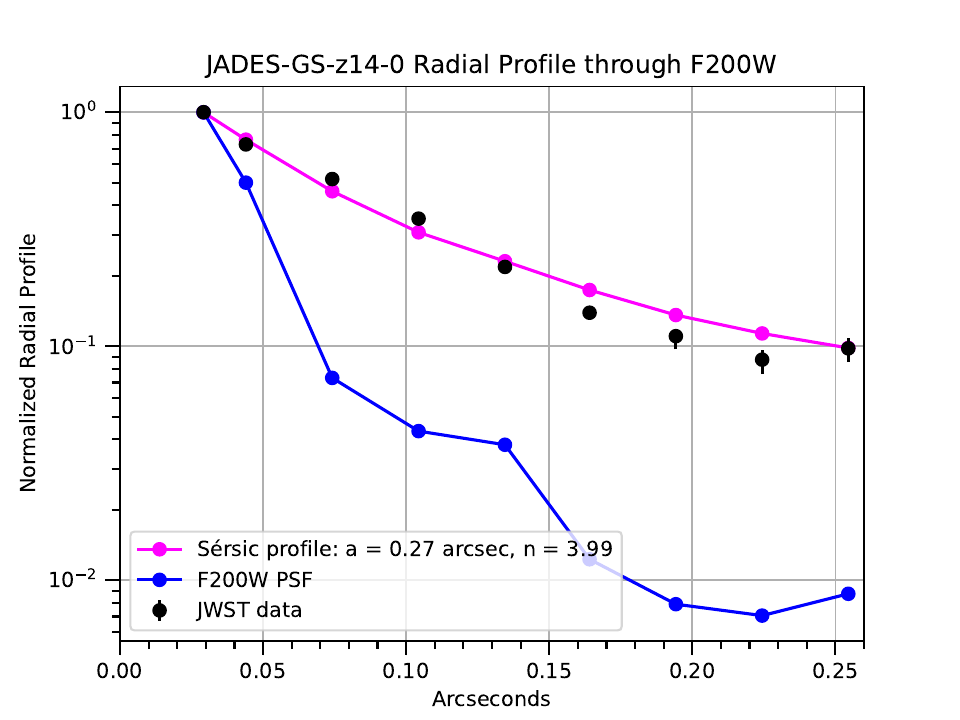}
    \caption{Morphology for \JADESfz}
    \label{fig:Morph_f}
    \end{subfigure}
\caption{Top panels: Corner plots representing 2D and 1D marginalized posteriors for the free parameters (S\'{e}rsic index and angular size in arcseconds) of our H ionization bound nebulae powered by Supermassive Dark Star models for \JADESeleven, \JADESzthirteen, and \JADESfz. For more details on the set up of the Monte Carlo simulation see Methods. Bottom panels: Radial profiles though F200W NIRCam band for each object. Data is represented by black dots, whereas our best fit model is shown in red. The blue line (and dots) represent the Point Spread Function (PSF) for F200W.}
\label{fig:morphology}
\end{figure*}

{\bf Morphology fits:} \JADESfo is unresolved~\citep{carniani2024shining}. As such, its morphology matches that of a point source, such as an isolated Supermassive Dark Star. As explained in our Methods section, the other three JADES objects considered are compact, yet extended~\citep{hainline2024searching,carniani2024shining}. We model them as SMDSs powering a hydrogen ionization bound nebula, as described in detail in the Methods section. In Fig.~\ref{fig:morphology} we present our results of the MC simulation for fitting the F200W radial profiles of \JADESeleven, \JADESzthirteen, and \JADESfz with the Supermassive Dark Stars identified by the spectral fits already described.  We restricted our analysis to a spherically symmetric nebula, and as such biased the S\'{e}rsic index to be sampled from a Gaussian distribution centered around $n=4$. Even with this restrictive assumption we find that the morphology of each of those three objects  is consistent with our Supermassive Dark Star models. In Table~\ref{tab:morphology} (pg.~\pageref{tab:morphology}) we summarize the values for our SMDS nebulae best fit parameters for the S\'{e}rsic index and angular size of each of the three resolved objects considered. 

\begin{table}
\centering
\caption{ Summary of best fit morphological parameters \label{tab:morphology}}
\begin{tabular}{lll}
\toprule
Object Name & Angular Size (arcsec) & S\'{e}rsic Index \\
\midrule
\JADESeleven &  $0.10^{+0.04}_{-0.04}$ & $4.00^{+0.05}_{-0.05}$\\
\JADESzthirteen & $0.066^{+0.0035}_{-0.0035}$ & $4.00^{+0.03}_{-0.03}$ \\
\JADESfz & $0.27^{+0.01}_{-0.01}$ & $3.99^{+0.04}_{-0.04}$ \\

\bottomrule
\end{tabular}
\end{table}

{\bf Spectral Features of Dark Stars:} In view of their relatively cool surface temperatures and significant amount of singly ionized helium (i.e. He~II) in their atmospheres, Dark Stars have a distinct smoking gun signature: an absorption feature due to He~II at 1640~{\AA}. No other known high redshift objects are expected to produce such an absorption feature. Early galaxies, with significant nebular emission, might show at the same wavelength an emission feature. Thus, detection of the He~II$\lambda1640$ absorption feature would imply a conclusive evidence for a  existence of those exotic objects. 
Further, the discovery of Dark stars could herald the first discovery of Dark Matter as well; further studies could then elucidate the mass and interaction strength of
the DM particles. 

\begin{figure}[!htb]
    \centering
    \includegraphics[width=0.9\linewidth]{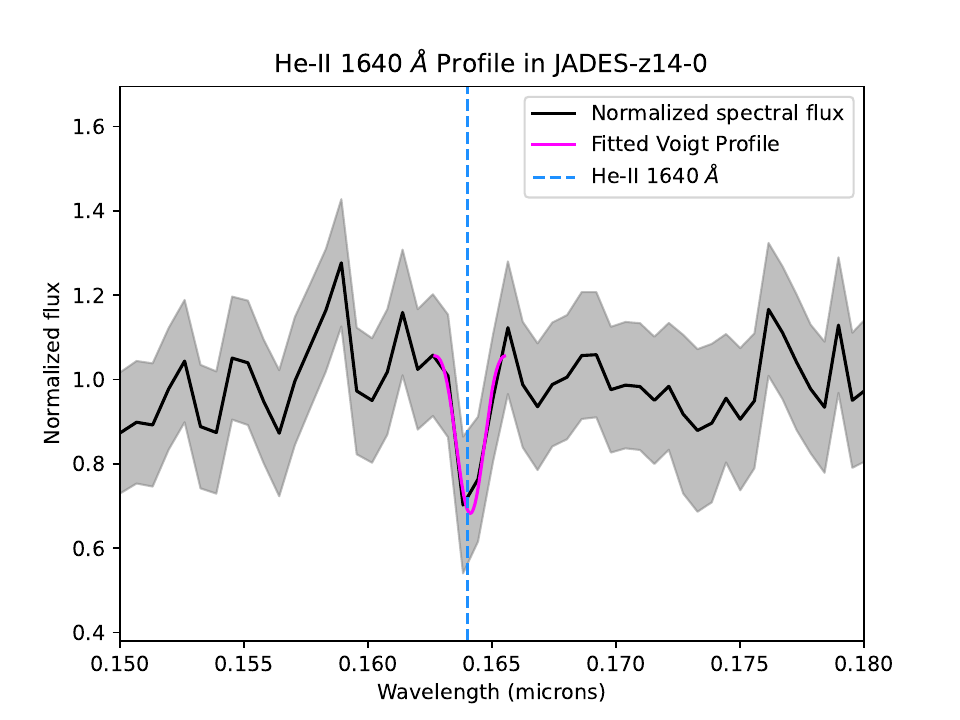}
    \caption{Equivalent width of the potential He II absorption feature in JADES-GS-z14-0. The normalized spectral flux is plotted in black, the Voigt profile fit to the feature is in pink, and the line center is shown in the dashed blue line. The uncertainty in the flux is shown in gray. The equivalent width is calculated based on the Voigt profile fit, and found to be 3.26 $\AA$. Note that given the extremely high error, followup spectroscopy is required to confirm the existence of this feature. Assumed $z=14.52$}
    \label{fig:jz14_he1640_ew}
\end{figure}

In Fig.~\ref{fig:jz14_he1640_ew} we show the identification of a He~II$\lambda1640$ absorption feature (S/N$\simeq 2.37$) in the NIRSpec observed flux of \JADESfz (the most distant object ever observed). The spectroscopic redshift corresponding to this feature is $z\simeq 14.52$, which lies very close to the $2-\sigma$ confidence level of the best fit redshift we found in our global NIRSpec SED Dark Star fit (see Table~\ref{tab:jades_properties}).~\footnote{We furthermore performed a separate MC parameter fit, where we fixed $z=14.52$ and let only the Dark Star mass and $n_H$ run. This new fit shows results consistent with our previous runs, as neither the stellar mas nor $n_H$ change appreciably.} While encouraging and very exciting, we caution the reader not to interpret this as conclusive evidence for the existence of Dark Stars. First and foremost, the Signal to Noise ratio is small (i.e. $S/N \sim 2$). Moreover, Ref.~\cite{carniani2025eventful} reports the discovery, with ALMA, of a [O~III]88$\mu$m emission line ($S/N~\sim6$) in the spectrum of \JADESfz, indicating a high level of metal enrichment in this system, not compatible with isolated Dark Stars.
At the end of the Discussion and Conclusion Section, we discuss the implications of these spectral lines if confirmed in followup observations including the possibility of Dark Stars in metal enriched environments.

\section*{Discussion and Conclusions} 

In this paper we used publicly available NIRSpec JWST data for four high redshift spectroscopically confirmed objects previously found by the JADES survey in order to test the Dark Star hypothesis. Three of those objects (\JADESeleven, \JADESzthirteen, and \JADESfz) are compact, yet resolved with NIRCam. We modeled them as Dark Stars powering a zero metallicity spherical nebula, with H and He in primordial abundance. \JADESfo is unresolved, and as such we model it as a pure supermassive Dark Star. We find compelling Dark Star fits for the observed spectra (see Fig.~\ref{fig:SEDFits}) and morphology (see Fig.~\ref{fig:morphology}) for each object analyzed. Furthermore, for \JADESfz, we identify a tentative He~II $\lambda1640$ absorption feature (see Fig.~\ref{fig:jz14_he1640_ew}). As discussed above, caution is warranted when interpreting this potential feature as evidence for the Dark Star hypothesis, especially in view of the relatively low S/N (2.37) and the very likely presence of a [O~III]88$\mu$m emission line from this system~\citep{carniani2025eventful}. Only followup observation would conclusively rule out or confirm the Dark Star hypothesis for this object, which is the first one for which we ever found evidence of a possible He~II absorption feature. Both features (i.e. O~III emission and He~II absorption) surviving  followup observations would be an extremely interesting scenario, as it would open up many questions. How could a Dark Star find itself in such a metal rich environment? Could it have formed there, or perhaps are mergers at play? While we don't know the answers to any of those questions yet, we foresee that they will not be specific to \JADESfz alone. As such, we plan to investigate them at a theoretical level in the near future.

\acknow{C.I. acknowledges funding from Colgate University via the Research Council and the Picker Interdisciplinary Science Institute. K. F. is grateful for support from the
Jeff \& Gail Kodosky Endowed Chair in Physics  at the University of Texas at Austin. K. F.  also acknowledges funding from the US Department
of Energy under Grant DE-SC-0022021, as well as the
Swedish Research Council (Contract No. 638-2013-8993).}

\showacknow{} 



\bibliography{ReferencesDMPopIII}

\begin{thebibliography}{10}

\bibitem{Abel:2001}
T Abel, GL Bryan, ML Norman, {The formation of the first star in the Universe}.
\newblock {\em\protect\JournalTitle{Science}} \textbf{295}, 93 (2002).

\bibitem{Barkana:2000}
R Barkana, A Loeb, {In the beginning: The First sources of light and the reionization of the Universe}.
\newblock {\em\protect\JournalTitle{Phys. Rept.}} \textbf{349}, 125--238 (2001).

\bibitem{Bromm:2003}
V Bromm, RB Larson, {The First stars}.
\newblock {\em\protect\JournalTitle{Ann. Rev. Astron. Astrophys.}} \textbf{42}, 79--118 (2004).

\bibitem{Yoshida:2006}
N Yoshida, K Omukai, L Hernquist, T Abel, {Formation of Primordial Stars in a lambda-CDM Universe}.
\newblock {\em\protect\JournalTitle{Astrophys. J.}} \textbf{652}, 6--25 (2006).

\bibitem{OShea:2007}
BW O'Shea, ML Norman, {Population III star formation in a lambda-CDM Universe. 1. The effect of formation redshift and environment on protostellar accretion rate}.
\newblock {\em\protect\JournalTitle{Astrophys. J.}} \textbf{654}, 66--92 (2007).

\bibitem{Yoshida:2008}
N Yoshida, K Omukai, L Hernquist, {Protostar Formation in the Early Universe}.
\newblock {\em\protect\JournalTitle{Science}} \textbf{321}, 669 (2008).

\bibitem{Bromm:2009}
V {Bromm}, N {Yoshida}, L {Hernquist}, CF {McKee}, {The formation of the first stars and galaxies}.
\newblock {\em\protect\JournalTitle{Nature}} \textbf{459}, 49--54 (2009).

\bibitem{Spolyar:2008dark}
D Spolyar, K Freese, P Gondolo, {Dark matter and the first stars: a new phase of stellar evolution}.
\newblock {\em\protect\JournalTitle{Phys. Rev. Lett.}} \textbf{100}, 051101 (2008).

\bibitem{Freese:2008ds}
K Freese, P Bodenheimer, D Spolyar, P Gondolo, {Stellar Structure of Dark Stars: a first phase of Stellar Evolution due to Dark Matter Annihilation}.
\newblock {\em\protect\JournalTitle{Astrophys. J.}} \textbf{685}, L101--L112 (2008).

\bibitem{Iocco:2008}
F Iocco, et~al., {Dark matter annihilation effects on the first stars}.
\newblock {\em\protect\JournalTitle{Monthly Notices of the Royal Astronomical Society}} \textbf{390}, 1655--1669 (2008).

\bibitem{Wu:2022wzw}
Y {Wu}, S {Baum}, K {Freese}, L {Visinelli}, HB {Yu}, {Dark Stars Powered by Self-Interacting Dark Matter}.
\newblock {\em\protect\JournalTitle{arXiv e-prints}} p. arXiv:2205.10904 (2022).

\bibitem{Freese:2010smds}
K Freese, C Ilie, D Spolyar, M Valluri, P Bodenheimer, {Supermassive Dark Stars: Detectable in JWST}.
\newblock {\em\protect\JournalTitle{Astrophys. J.}} \textbf{716}, 1397--1407 (2010).

\bibitem{Zackrisson:2010HighZDS}
E {Zackrisson}, et~al., {Finding High-redshift Dark Stars with the James Webb Space Telescope}.
\newblock {\em\protect\JournalTitle{The Astrophysical Journal}} \textbf{717}, 257--267 (2010).

\bibitem{Ilie:2012}
C {Ilie}, K {Freese}, M {Valluri}, IT {Iliev}, PR {Shapiro}, {Observing supermassive dark stars with James Webb Space Telescope}.
\newblock {\em\protect\JournalTitle{Monthly Notices of the Royal Astronomical Society}} \textbf{422}, 2164--2186 (2012).

\bibitem{Zhang:2022}
S Zhang, C Ilie, K Freese, {Detectabilty of Supermassive Dark Stars with the Roman Space Telescope}.
\newblock {\em\protect\JournalTitle{arXiv}} (2023).

\bibitem{Ilie:2023JADES}
C Ilie, J Paulin, K Freese, {Supermassive Dark Star candidates seen by JWST}.
\newblock {\em\protect\JournalTitle{Proc. Nat. Acad. Sci.}} \textbf{120}, e2305762120 (2023).

\bibitem{hubeny2017tlusty}
I Hubeny, T Lanz, Tlusty user's guide iii: Operational manual.
\newblock {\em\protect\JournalTitle{arXiv preprint arXiv:1706.01937}} (2017).

\bibitem{carniani2025eventful}
S Carniani, et~al., The eventful life of a luminous galaxy at z= 14: metal enrichment, feedback, and low gas fraction?
\newblock {\em\protect\JournalTitle{Astronomy \& Astrophysics}} \textbf{696}, A87 (2025).

\bibitem{GLASSz13}
RP {Naidu}, et~al., {Two Remarkably Luminous Galaxy Candidates at $z\approx11-13$ Revealed by JWST}.
\newblock {\em\protect\JournalTitle{arXiv e-prints}} p. arXiv:2207.09434 (2022).

\bibitem{Maisies:2022}
SL {Finkelstein}, et~al., {A Long Time Ago in a Galaxy Far, Far Away: A Candidate z \raisebox{-0.5ex}\textasciitilde 14 Galaxy in Early JWST CEERS Imaging}.
\newblock {\em\protect\JournalTitle{arXiv e-prints}} p. arXiv:2207.12474 (2022).

\bibitem{z16.CEERS93316:2022}
CT {Donnan}, et~al., {The evolution of the galaxy UV luminosity function at redshifts z \raisebox{-0.5ex}\textasciitilde 8-15 from deep JWST and ground-based near-infrared imaging}.
\newblock {\em\protect\JournalTitle{arXiv e-prints}} p. arXiv:2207.12356 (2022).

\bibitem{z17.Schrodinger:2022}
RP {Naidu}, et~al., {Schrodinger's Galaxy Candidate: Puzzlingly Luminous at $z\approx17$, or Dusty/Quenched at $z\approx5$?}
\newblock {\em\protect\JournalTitle{arXiv e-prints}} p. arXiv:2208.02794 (2022).

\bibitem{JADES:2022a}
BE {Robertson}, et~al., {Discovery and properties of the earliest galaxies with confirmed distances}.
\newblock {\em\protect\JournalTitle{arXiv e-prints}} p. arXiv:2212.04480 (2022).

\bibitem{JADES:2022b}
E {Curtis-Lake}, et~al., {Spectroscopic confirmation of four metal-poor galaxies at z=10.3-13.2}.
\newblock {\em\protect\JournalTitle{arXiv e-prints}} p. arXiv:2212.04568 (2022).

\bibitem{Labbe:2022}
I {Labbe}, et~al., {A population of red candidate massive galaxies \raisebox{-0.5ex}\textasciitilde600 Myr after the Big Bang}.
\newblock {\em\protect\JournalTitle{arXiv e-prints}} p. arXiv:2207.12446 (2022).

\bibitem{RedMonsters:2024}
M Xiao, et~al., Accelerated formation of ultra-massive galaxies in the first billion years.
\newblock {\em\protect\JournalTitle{Nature}} \textbf{635}, 311--315 (2024).

\bibitem{Finkelstein_2023}
SL Finkelstein, et~al., Ceers key paper. i. an early look into the first 500 myr of galaxy formation with jwst.
\newblock {\em\protect\JournalTitle{The Astrophysical Journal Letters}} \textbf{946}, L13 (2023).

\bibitem{harikane2023purespectroscopicconstraintsuv}
Y Harikane, et~al., Pure spectroscopic constraints on uv luminosity functions and cosmic star formation history from 25 galaxies at $z_\mathrm{spec}=8.61-13.20$ confirmed with jwst/nirspec (2023).

\bibitem{Yung:2024}
LYA Yung, RS Somerville, SL Finkelstein, SM Wilkins, JP Gardner, Are the ultra-high-redshift galaxies at z \&gt; 10 surprising in the context of standard galaxy formation models?
\newblock {\em\protect\JournalTitle{Monthly Notices of the Royal Astronomical Society}} \textbf{527}, 5929--5948 (2023).

\bibitem{JZ:2025}
J Ziegler, J Lozano~Mayo, G Montefalcone, K Freese, In preparation (2025).

\bibitem{Dekel_2023}
A Dekel, KC Sarkar, Y Birnboim, N Mandelker, Z Li, Efficient formation of massive galaxies at cosmic dawn by feedback-free starbursts.
\newblock {\em\protect\JournalTitle{Monthly Notices of the Royal Astronomical Society}} \textbf{523}, 3201–3218 (2023).

\bibitem{BHdorm24}
I Juod{\v{z}}balis, et~al., A dormant overmassive black hole in the early universe.
\newblock {\em\protect\JournalTitle{Nature}} \textbf{636}, 594--597 (2024).

\bibitem{Bogdan:2023UHZ1}
{\'A} {Bogd{\'a}n}, et~al., {Evidence for heavy-seed origin of early supermassive black holes from a z {\ensuremath{\approx}} 10 X-ray quasar}.
\newblock {\em\protect\JournalTitle{Nature Astronomy}} \textbf{8}, 126--133 (2024).

\bibitem{Wang:2019}
F {Wang}, et~al., {Exploring Reionization-era Quasars. III. Discovery of 16 Quasars at 6.4 {\ensuremath{\lesssim}} z {\ensuremath{\lesssim}} 6.9 with DESI Legacy Imaging Surveys and the UKIRT Hemisphere Survey and Quasar Luminosity Function at z {\ensuremath{\sim}} 6.7}.
\newblock {\em\protect\JournalTitle{The Astrophysical Journal}} \textbf{884}, 30 (2019).

\bibitem{Inayoshi:2020}
K Inayoshi, E Visbal, Z Haiman, The assembly of the first massive black holes.
\newblock {\em\protect\JournalTitle{Annual Review of Astronomy and Astrophysics}} \textbf{58}, 27--97 (2020).

\bibitem{Lupi:2021}
A Lupi, Z Haiman, M Volonteri, {Forming massive seed black holes in high-redshift quasar host progenitors}.
\newblock {\em\protect\JournalTitle{Mon. Not. Roy. Astron. Soc.}} \textbf{503}, 5046--5060 (2021).

\bibitem{ilie2023uhz1}
C Ilie, K Freese, A Petric, J Paulin, Uhz1 and the other three most distant quasars observed: possible evidence for supermassive dark stars.
\newblock {\em\protect\JournalTitle{arXiv preprint arXiv:2312.13837}} (2023).

\bibitem{Loeb:1994wv}
A Loeb, FA Rasio, {Collapse of primordial gas clouds and the formation of quasar black holes}.
\newblock {\em\protect\JournalTitle{Astrophys. J.}} \textbf{432}, 52 (1994).

\bibitem{Belgman:2006}
MC {Begelman}, M {Volonteri}, MJ {Rees}, {Formation of supermassive black holes by direct collapse in pre-galactic haloes}.
\newblock {\em\protect\JournalTitle{Monthly Notices of the Royal Astronomical Society}} \textbf{370}, 289--298 (2006).

\bibitem{Lodato:2006hw}
G Lodato, P Natarajan, {Supermassive black hole formation during the assembly of pre-galactic discs}.
\newblock {\em\protect\JournalTitle{Mon. Not. Roy. Astron. Soc.}} \textbf{371}, 1813--1823 (2006).

\bibitem{Natarajan:2017}
P Natarajan, , et~al., {Mapping substructure in the HST Frontier Fields cluster lenses and in cosmological simulations}.
\newblock {\em\protect\JournalTitle{Mon. Not. Roy. Astron. Soc.}} \textbf{468}, 1962--1980 (2017).

\bibitem{barrow:2018}
KSS Barrow, A Aykutalp, JH Wise, Observational signatures of massive black hole formation in the early universe.
\newblock {\em\protect\JournalTitle{Nature Astronomy}} \textbf{2}, 987–994 (2018).

\bibitem{Whalen:2020}
DJ Whalen, et~al., Finding direct-collapse black holes at birth.
\newblock {\em\protect\JournalTitle{The Astrophysical Journal Letters}} \textbf{897}, L16 (2020).

\bibitem{Banik:2019}
N {Banik}, JC {Tan}, P {Monaco}, {The formation of supermassive black holes from Population III.1 seeds. I. Cosmic formation histories and clustering properties}.
\newblock {\em\protect\JournalTitle{Monthly Notices of the Royal Astronomical Society}} \textbf{483}, 3592--3606 (2019).

\bibitem{Singh_2023}
J Singh, P Monaco, JC Tan, The formation of supermassive black holes from population iii.1 seeds. ii. evolution to the local universe.
\newblock {\em\protect\JournalTitle{Monthly Notices of the Royal Astronomical Society}} \textbf{525}, 969–982 (2023).

\bibitem{cammelli2024formationsupermassiveblackholes}
V Cammelli, et~al., The formation of supermassive black holes from population iii.1 seeds. iii. galaxy evolution and black hole growth from semi-analytic modelling (2024).

\bibitem{Loeb:2010}
A Loeb, {\em {How did the first stars and galaxies form?}}
\newblock (Princeton University Press, Princeton, NJ), (2010).

\bibitem{Bromm:2011}
V Bromm, N Yoshida, The first galaxies.
\newblock {\em\protect\JournalTitle{Annual Review of Astronomy and Astrophysics}} \textbf{49}, 373--407 (2011).

\bibitem{Gnedin:2016}
NY Gnedin, Cosmic reionization on computers: the faint end of the galaxy luminosity function.
\newblock {\em\protect\JournalTitle{The Astrophysical Journal Letters}} \textbf{825}, L17 (2016).

\bibitem{Dayal:2018}
P Dayal, A Ferrara, Early galaxy formation and its large-scale effects.
\newblock {\em\protect\JournalTitle{Physics Reports}} \textbf{780}, 1--64 (2018).

\bibitem{Yung:2019}
LA Yung, RS Somerville, SL Finkelstein, G Popping, R Dav{\'e}, Semi-analytic forecasts for jwst--i. uv luminosity functions at z= 4--10.
\newblock {\em\protect\JournalTitle{Monthly Notices of the Royal Astronomical Society}} \textbf{483}, 2983--3006 (2019).

\bibitem{Behroozi:2020}
P Behroozi, et~al., The universe at z> 10: predictions for jwst from the universemachine dr1.
\newblock {\em\protect\JournalTitle{Monthly Notices of the Royal Astronomical Society}} \textbf{499}, 5702--5718 (2020).

\bibitem{Hirano:2014}
S Hirano, et~al., {ONE} {HUNDRED} {FIRST} {STARS}: {PROTOSTELLAR} {EVOLUTION} {AND} {THE} {FINAL} {MASSES}.
\newblock {\em\protect\JournalTitle{The Astrophysical Journal}} \textbf{781}, 60 (2014).

\bibitem{Spolyar:2009}
D Spolyar, P Bodenheimer, K Freese, P Gondolo, {Dark Stars: a new look at the First Stars in the Universe}.
\newblock {\em\protect\JournalTitle{Astrophys. J.}} \textbf{705}, 1031--1042 (2009).

\bibitem{Freese:2016dark}
K Freese, T Rindler-Daller, D Spolyar, M Valluri, Dark stars: a review.
\newblock {\em\protect\JournalTitle{Reports on Progress in Physics}} \textbf{79}, 066902 (2016).

\bibitem{Blumenthal:1985}
GR Blumenthal, S Faber, R Flores, JR Primack, Contraction of dark matter galactic halos due to baryonic infall.
\newblock {\em\protect\JournalTitle{The Astrophysical Journal}} \textbf{301}, 27--34 (1986).

\bibitem{Freese:2008dmdens}
K Freese, P Gondolo, J Sellwood, D Spolyar, {Dark Matter Densities during the Formation of the First Stars and in Dark Stars}.
\newblock {\em\protect\JournalTitle{Astrophys. J.}} \textbf{693}, 1563--1569 (2009).

\bibitem{Paxton2011}
B {Paxton}, et~al., {Modules for Experiments in Stellar Astrophysics (MESA)}.
\newblock {\em\protect\JournalTitle{The Astrophysical Journals}} \textbf{192}, 3 (2011).

\bibitem{Rindler-Daller:2015SMDS}
T {Rindler-Daller}, MH {Montgomery}, K {Freese}, DE {Winget}, B {Paxton}, {Dark Stars: Improved Models and First Pulsation Results}.
\newblock {\em\protect\JournalTitle{The Astrophysical Journal}} \textbf{799}, 210 (2015).

\bibitem{chatzikos20232023}
M Chatzikos, et~al., The 2023 release of cloudy.
\newblock {\em\protect\JournalTitle{arXiv preprint arXiv:2308.06396}} (2023).

\bibitem{Zackrisson:2011}
E {Zackrisson}, {The observational signatures of high-redshift dark stars}.
\newblock {\em\protect\JournalTitle{arXiv e-prints}} p. arXiv:1101.2895 (2011).

\bibitem{GS12:2024jades}
F d’Eugenio, et~al., Jades: Carbon enrichment 350 myr after the big bang.
\newblock {\em\protect\JournalTitle{Astronomy \& Astrophysics}} \textbf{689}, A152 (2024).

\bibitem{hainline2024searching}
KN Hainline, et~al., Searching for emission lines at z> 11: The role of damped ly$\alpha$ and hints about the escape of ionizing photons.
\newblock {\em\protect\JournalTitle{The Astrophysical Journal}} \textbf{976}, 160 (2024).

\bibitem{carniani2024shining}
S Carniani, et~al., Spectroscopic confirmation of two luminous galaxies at a redshift of 14.
\newblock {\em\protect\JournalTitle{Nature}} \textbf{633}, 318–322 (2024).

\bibitem{Hubeney:1988}
I {Hubeny}, {A computer program for calculating non-LTE model stellar atmospheres}.
\newblock {\em\protect\JournalTitle{Computer Physics Communications}} \textbf{52}, 103--132 (1988).

\bibitem{Pickering_2016}
TE Pickering, et~al., Pandeia: a multi-mission exposure time calculator for jwst and wfirst in {\em Observatory Operations: Strategies, Processes, and Systems VI}, eds.{} AB Peck, CR Benn, RL Seaman.
\newblock (SPIE), (2016).

\bibitem{Sersic:1968}
JL {Sersic}, {\em {Atlas de Galaxias Australes}}.
\newblock (1968).

\bibitem{Gunn-Peterson:1965}
JE {Gunn}, BA {Peterson}, {On the Density of Neutral Hydrogen in Intergalactic Space.}
\newblock {\em\protect\JournalTitle{The Astrophysical Journal}} \textbf{142}, 1633--1636 (1965).

\end{thebibliography}

\end{document}